\documentclass[11pt]{article}
\usepackage{geometry}                % See geometry.pdf to learn the layout options. There are lots.
\usepackage{graphicx}
\usepackage{amssymb}
\usepackage{epstopdf}
\title{Electrostatic map of T7 DNA. Comparative analysis of functional and electrostatic properties of T7 RNA polymerase specific promoters.}
\author{S.G.Kamzolova$^{1}$, P.M.Beskaravainy$^{1,2}$, A.A.Osypov$^{1}$, T.R.Dzhelyadin$^{1}$, E.A.Temlyakova$^{1}$ \\
        and A.A.Sorokin$^{1,3} $\\
        \\
	~$^{1}$Institute of Cell Biophysics RAS,  Pushchino,  Russia \\
	~$^{2}$Institute of Theoretical and Experimental Biophysics RAS, Pushchino,Russia\\
	~$^{3}$School of Informatics, The University of Edinburgh, Edinburgh, UK\\
	}

\begin{document}

\maketitle
\begin{abstract}
The entire T7 bacteriophage genome contains 39937 base pairs (Database NCBI RefSeq N1001604). Here, electrostatic potential distribution around double helical T7 DNA was calculated by Coulomb method. Electrostatic profiles of 17 promoters recognized by T7 phage specific RNA polymerase were analyzed. It was shown that electrostatic profiles of all T7 RNA polymerase specific promoters can be characterized by distinctive motifs which are specific for each promoter class. Comparative analysis of electrostatic profiles of native T7 promoters of different classes demonstrates that  T7 RNA polymerase can differentiate them due to their electrostatic features. Software to calculate distribution of electrostatic potentials is available from http://promodel.github.io/reldna/.
\end{abstract}
%\subsection{}
\section{Introduction}
The problem of RNA polymerase-promoter recognition has a long and abundant history. A more traditional line of investigation considers nucleotide sequence elements of promoter structure as the only recognizable components that are important. Promoter search algorithms of this type are mainly based on the sequence preferences in the regions of specific contacts with RNA polymerase (Alexandrov \& Mironov, 1990; Gordon, Chervonenkis, Gammerman, Shahmuradov, \& Solovyev, 2003; Hertz \& Stormo, 1996; Horton \& Kanehisa, 1992; Huerta \& Collado-Vides, 2003; Mahadevan \& Ghosh, 1994; Pedersen \& Engelbrecht, 1995; Vanet, Marsan, \& Sagot, 1999; Yada, Nakao, Totoki, \& Nakai, 1999). Most of them can identify 80-90\% promoters from tested compilations but the principal shortcoming of such algorithms lies in recognition of many false signals. Even the best protocols at this level recognize a large portion of nonpromoter DNA (up to 3,5\%) as promoter like signals (Gordon et al., 2003; Horton \& Kanehisa, 1992). Within \textit{Escherihia coli}genome, the background noise is, therefore, more than one order of magnitude greater than the genuine signal.

Modern research tends to call attention to some additional information encoded in physical properties of DNA helix. Some physicochemical characteristics of promoter DNA, such as overall geometry, deformability, thermal instability and dynamical features were shown to be such discriminating factors involved in differential interaction of RNA polymerase with various promoters (deHaseth \& Helmann, 1995; Kamzolova, Ivanova, \& Kamzalov, 1999; Kamzolova \& Postnikova, 1981; Leirmo \& Gourse, 1991; Margalit, Shapiro, Nussinov, Owens, \& Jernigan, 1988; Pérez-Martín, Rojo, \& de Lorenzo, 1994; Travers, 1989).

We approached the problem from the analysis of electrostatic properties of promoter DNA (Kamzolova et al., 2000; Polozov et al., 1999). Our study of electrostatics contribution in RNA polymerase-promoter recognition was made possible due to development of an original method of electrostatic potential calculations around DNA helix suitable for a genome wide application (Polozov et al., 1999; Sorokin, 2001). The method takes advantage of observing electrostatic profiles of promoters within the electrostatic map of a whole genome DNA. Using this method, the results were obtained indicating essential differences in electrostatic patterns for promoter and nonpromoter DNA in ~\textit{E.~coli} genome (Kamzolova, Sorokin, Beskaravainy, \& Osypov, 2006; Kamzolova, Sorokin, Dzhelyadin, Beskaravainy, \& Osypov, 2005; Polozov et al., 1999). Some characteristic electrostatic elements specified as new promoter determinants were found in upstream region of σ70-specific promoters in genomes of \textit{Escherihia coli} and its related bactriophages (T4, T7 or $\lambda$) (Kamzolova et al., 2005; Kamzolova, Sorokin, Osipov, \& Beskaravainy, 2009; Osypov, Krutinin, \& Kamzolova, 2010; Sorokin, Osypov, Dzhelyadin, Beskaravainy, \& Kamzolova, 2006; Sorokin, 2001). Moreover, the role of DNA electrostatic potential in promoter organization was confirmed by the fact that, in \textit{Escherihia coli}, the most negatively charged oligonucleotides were revealed in promoter regions as compared with the total genome structure (Sorokin, Osipov, Beskaravainy, \& Kamzolova, 2007; Sorokin, 2001). It also should be mentioned that relationship between promoter sequence and its electrostatic profile was found to be ambiguous (Kamzolova, Osypov, Dzhelyadin, Beskaravainy, \& Sorokin, 2006; Kamzolova, Sorokin, et al., 2006; Sorokin, Dzhelyadin, Ivanova, Polozov, \& Kamzolova, 2001; Sorokin, 2001), meaning that this property is vastly dependent on the whole sequence with flanking regions rather than the sequence text at the given point of consideration. So, we must consider electrostatic property as a promoter determinant in its own right thus, emphasizing the need for studying electrostatic characteristics of a promoter in addition to its text analysis.

In this paper distribution of electrostatic potential around the complete sequence of T7 phage genome was calculated. In contrast to the studies above carried out with ~\textit{E.~coli} RNA polymerase specific promoters, here, electrostatic properties of T7 promoters recognized by T7 phage RNA polymerase were studied. T7 specific promoters are shown to differ from nonpromoter DNA and σ70-specific promoters by their electrostatic characteristics. T7 specific promoters can be groupped into two classes by their biochemical and physiological properties. The promoters assigned to the different classes are found to differ also by their electrostatic properties. Thus, the results obtained indicate that T7 RNA polymerase can identify its native promoters in T7 genome and differentiate them due to their electrostatic features.

\section{Methods}
Nucleotide sequence of T7 phage genome and its annotation involving localization of promoters, terminators and identified genes were taken from NCBIRefSeq Database (ftp://ftp.ncbi.nih.gov/refseq). The electrostatic potential around the double helical DNA molecule was calculated by the original method based on the Coulomb's law (Kamzolova et al., 2000; Polozov et al., 1999) using the computer program of A.A.Sorokin applicable for the whole genome calculation (Sorokin, 2001) and available http://promodel. github.io/reldna/. Full details of the method is presented in Appendix.

Full-atomic 3D model of the T7 RNA polymerase was obtained from SwissModel (Kiefer, et al. 2009) by UniProt ID P00573 and was equilibrated at 300K in water at 100 mM salt for 5 ns with OPLS force field (Jorgensen and Tirado-Rives 1988) in GROMOS software (Hess, et al. 2008). Electrostatic potential distribution around protein globule was calculated with APBS software (Baker et al. 2001) at 310K, neutral pH and100 mM of univalent salt.

\section{Results and discussion}
\subsection{T7 phage genome. Classification of T7 late genes and promoters. Electrostatic properties of T7 promoter and nonpromoter DNA.}
T7 phage genome consists of 39937 b.p. containing information for more than 50 proteins. Temporal coordination of T7 DNA expression is controlled by two RNA polymerases (Dunn, Studier, \& Gottesman, 1983). The early region of the genome located in its left part (~20\% of DNA) is transcribed by the host RNA polymerase (Eσ70) immediately after infection with resulting efficient production of T7 RNA polymerase during 6 min. The remaining part of the genome containing late genes is transcribed by the newly synthesized T7 RNA polymerase.
The late genes may be classified in two groups (class II and class III) by their location in the genome and their expression time (Dunn et al., 1983). Transcription of class II genes which encode proteins involved in the phage DNA metabolism takes place in the time interval from 6 to 15 min after infection. Genes of class III containing information for viral structural proteins are transcribed from 8 min after infection till cell lysis. There are 17 T7 RNA polymerase specific promoters controlling expression of the late genes in T7 DNA. They can also be classified according to the categories of the corresponding genes. The main characteristics of the promoters (site location in the genome, category, nucleotide sequence) are given in Table 1.

10 promoters fall in the category of class II. However, it should be mentioned that the terminator TE1 for E.coli RNA polymerase lies in T7 genome at the position of 7588 b.p. downstream of three promoters from class II, namely φ1.1A, φ1.1B and φ1.3 which are located at the positions of 5848 b.p., 5923 b.p. and 6409 b.p. respectively. So, they fall within the coding region of T7 DNA transcribed by E.coli RNA polymerase. Their transcripts appear earlier as compared with other II class transcripts due to the uninterrupted RNA synthesis produced by E.coli RNA polymerase up to TE1 site. Therefore, these promoters form a separate group of «early» promoters (subgroup of II class promoters). Their location within DNA coding region can influence their physical properties.

5 promoters (φ6.5, φ9, φ10, φ13 and φ17) located in the right part of the genome represent class III. Two promoters, φOL and φOR are a special case because they control expression of no proteins. They are possibly involved in the process of DNA replication initiation. The φOL is located in the early region of the genome and can be formally assigned to the group of «early» promoters. The promoter φOR can formally be placed in class III according to its expression time.

Nucleotide sequences of all the promoters are known. As shown in Table 1, all promoters of class III have identical 23-membered nucleotide sequence from +6 b.p. to -17 b.p. This conservative sequence is considered as consensus sequence for T7 RNA polymerase. The other promoters can be also characterized by a high level of homology with the consensus sequence differing from it by 1-7 b.p. in the individual promoters.

Taking into account a small size of T7 RNA polymerase, this sequence is sufficiently large to be the only region involved in the formation of specific contacts with the enzyme thus indicating the importance of this area for RNA polymerase-promoter recognition in this case.
The promoters have been earlier characterized by their complexformation with RNA polymerase and transcription initiation (McAllister \& Carter, 1980; McAllister, Morris, Rosenberg, \& Studier, 1981; McAllister \& Wu, 1978). Despite their considerable sequence similarity, the promoters of II and III classes essentially differ by their strengths and by some biochemical properties (Golomb \& Chamberlin, 1974; Kassavetis \& Chamberlin, 1979; McAllister \& Carter, 1980; McAllister et al., 1981; McAllister \& Wu, 1978; McAllister \& McCarron, 1977; Niles \& Condit, 1975; Studier \& Rosenberg, 1981), thus suggesting some contribution of physical properties in their functional behaviour.

Calculation of distribution of electrostatic potential around the complete sequence of T7 DNA was carried out. Electrostatic patterns of DNA fragments of 60 b.p. in length containing 17 individual late promoters (regions from -30 b.p. to +30 b.p. around transcription start point) and 17 nonpromoter randomly selected sites were chosen at T7 DNA electrostatic map and analyzed.

Electrostatic profiles of all promoters are shown in Fig. 1. Although electrostatic patterns of the individual promoters differ by their details, they have some common features. When superimposed, they reveal a well defined wave shaped design with a wide valley from +1 b.p. to -20 b.p. and a higher potential around the start point of transcription. By contrast, no common specific elements were found in electrostatic profiles of T7 DNA nonpromoter sites (Fig.2). Electrostatic pattern of superimposed profiles of 17 randomly selected DNA fragments can be characterized by a rather homogeneous variation of electrostatic potential around the mean potential value. The results are in agreement with those obtained for E.coli genome (Kamzolova, Sorokin, et al., 2006; Kamzolova et al., 2005; Polozov et al., 1999). Nonpromoter sites in the bacterial genome can be described by more homogeneous and smoothed electrostatic profiles as compared with complicated, rich in details patterns of σ70-specific promoters recognized by ~\textit{E.~coli} RNA polymerase. 

In addition, as evident from the analysis of Fig.1 and Fig.2, T7 promoter DNA as a whole is characterized by more negative potential values as compared with randomly selected sequences. This fact also agrees with the conclusion that the most negatively charged sequences are found just in promoter regions of ~\textit{E.~coli} genome DNA (Sorokin et al., 2007; Sorokin, 2001).

It should be mentioned that electrostatic profiles of promoters recognized by two RNA polymerases differ in their size and design. Characteristic electrostatic patterns for σ70- and T7-specific promoters embrace ~200 b.p. (-150 – +50) and ~40 b.p. (-25 – +15) correspondingly. The difference in the size of the patterns closely falls within the range of DNA contacting sites for the two enzymes what is to be accounted for by the difference in their size. ~\textit{E.~coli} RNA polymerase is a large multisubunit protein. One of its α-subunit forms specific contacts with upstream sites of promoter DNA. It is interesting that the most noticeable electrostatic signals involved in recognition of this enzyme were found in far upstream region of its native promoters (-70 – -120 b.p.) in various templates such as DNA of E.coli and related bacteriophages T4, T7 and λ (Kamzolova et al., 2000, 2005, 2009; Osypov et al., 2010; Sorokin et al., 2006). By contrast, there are no specific electrostatic elements in this region of T7 RNA polymerase specific promoters. All characteristic features of electrostatic profiles of these promoters are located in a small area containing the consensus region with some surrounding sequences (Fig. 2).

So, the results obtained indicate that electrostatic profiles of T7 specific promoters are marked with some peculiarities in the genome molecule thus favouring their identification by T7 RNA polymerase.

\subsection{Electrostatic properties of T7 RNA polymerase.}
Structure of T7 RNA polymerase that contains all parts of the protein including N-terminal 8 amino acids, which are generally missing in the crystal structures of the protein, were obtained from SwissModel. Comparison of the equvilibrated structure with original model shows that C-alpha atoms root mean square deviation was in a range 2-3Å. This deviation was mainly due to movement of loops (see also Supplementary Figure 1). Electrostatic potential around equilibrated model was calculated. It is characterized by great anisotropy (Supplementary Figure 2). The surface around active site and promoter-binding domain form a positively charged groove, while the opposite side of the protein are mainly negatively charged. Distribution of that kind could play a crucial role in the proper orientation of the protein towards the DNA during initial steps of promoter recognition: non-specific binding and promoter location.

It is known (Cheetham, Jeruzalmi, \& Steitz, 1999, Durniak, Bailey \& Steitz 2008) that there are two domains responsible for promoter binding during transcription initiation step: promoter-binding domain (PBD) formed by six a-helical bundle (residues 72-150 and 191-267) and specificity loop (residues 739-770). Two parts of PDB are enclosed specificity loop constituting monolithic promoter recognition domain (Fig. 3A). 
Distribution of electrostatic potential around that domain shows three clear crests of positive potential (Fig. 3B), which correspond to three sub-domains: inter-helical loop of the PBD (residues 72-150) responsible for binding with AT-rich sequence of promoter DNA around -17 bp (designated as site A at Fig. 3A); specificity loop binding elements around position -9 bp, mutation in which cause switching of specificity from the T7 to the T3 specific promoter sequences (site B at Fig. 3A) and the loop in the second part of PBD (residues 191-267) that interacts with promoter around transcriptional start site (site C at Fig. 3A).

Analysis of electrostatic potential distribution around three crests shows that the potential of the greatest positive value is located around PBD site A sub-domain, while PBD site C demonstrates more neutral values of the potential. Specificity loop (site B) is located between that two extrema. 

\subsection{Electrostatic properties of T7 RNA polymerase specific promoters of different classes.}
The comparative analysis of electrostatic properties of the promoters belonging to the different classes was carried out. Electrostatic profiles of the individual promoters of II and III classes are presented in Fig. 4 and Fig. 5 respectively. The averaged electrostatic profiles of these groups were calculated as indicated in (Kamzolova et al., 2009). The results representing the typical profiles for the each group are shown in Fig.6. 

Our attention was given to the analysis of the promoter region (-25 b.p. – +15 b.p.) which, as indicated above, is responsible for making contacts with the enzyme. The region in the both promoter classes can be characterized as an enormous negatively charged valley where T7 RNA polymerase could be seized and held in place to provide the next steps of promoter-polymerase conformational changes up to the production of the transcriptionally active complex. T7 RNA polymerase active sites involved in interaction with this region of the T7 promoter DNA are known (Cheetham, Jeruzalmi, \& Steitz, 1999). As shown in Fig. 3, these sites are mainly positively charged, thus supporting their electrostatic attraction by the corresponding promoter region (sites A, B and C).

T7 promoters were proposed to have at least three functional partially overlapping domains (Cheetham et al., 1999; Gunderson, Chapman, \& Burgess, 1987; Ikeda \& Richardson, 1987; Joho, Gross, McGraw, Raskin, \& McAllister, 1990; McAllister \& Carter, 1980; Muller, Martin, \& Coleman, 1989). One of them, so called “binding domain” (-17 b.p. – -6 b.p.) is involved in RNA polymerase identification and binding at early steps of their interaction. At these steps DNA electrostatic properties can contribute essentially to primary promoter-polymerase recognition (Kamzolova et al., 2000; Polozov et al., 1999; von Hippel, 2004).

Some T7 promoters (mainly of III class) have an additional functional AT-rich site in this domain (-23 b.p. – -13 b.p.) which is also involved in primary electrostatics dependent recognition of RNA polymerase.

The right part of T7 promoters is “initiation domain” (-5 b.p. – +6 b.p.) participating in the formation of the transcriptionally active complex at later steps when serious conformational changes involving DNA melting give rise to the “open complex” capable of transcription initiation in the presence of RNA synthesis substrates. It should be mentioned that electrostatic interactions are not involved in the formation of the “melted” complex. So, electrostatic properties of the “initiation domain” should have no essential significance for its interaction with the corresponding enzyme site at these stages of complex formation.

Thus, electrostatic properties of T7 promoter DNA can contribute differently to its complex with RNA polymerase at different stages of their interaction.

As shown in Figs. 4-6 two different characteristic electrostatic motifs are found in T7 promoter DNA. The first motif is formed in the “binding domain” (together with AT-rich site in the case of III group promoters). It represents a deep hole with the most negative potential value located around position -18 b.p. This is the most pronounced element in T7 promoters belonging to III class (Fig. 5 and Fig. 6). The element is engaged in primary identification of T7 promoter DNA by the most positively charged region of T7 RNA polymerase including AT-rich domain binding site A (in the case of III class promoters) and the site B (see Fig. 3). Electrostatic attraction between these functional elements at primary steps of their recognition should have a favourable effect on the complex thus indicating a beneficial contribution of DNA electrostatic characteristics to promoter activity in the case of III class promoters. It is known that T7 promoters of III class are “strong” promoters (Cheetham et al., 1999; Golomb \& Chamberlin, 1974; McAllister \& McCarron, 1977; Niles \& Condit, 1975). Taking into account the results obtained, it can be suggested that the high level of their activity will be provided not only by their perfect nucleotide sequence fully identical to the consensus sequence but by their appropriate electrostatic properties as well.

T7 promoters of II class are characterized by a distinctly different electrostatic profile (Fig. 4 and Fig. 6). They contain two different characteristic elements. Both elements are negatively charged holes located at ~ -18 b.p. and ~ -5 b.p. in a wide valley with the smooth rise of potential downstream of -18 b.p. and sharp increase of potential upstream of -5 b.p. to the maximum at ~+5 b.p. The first element is located in the “binding domain” of promoter DNA. It is similar to the motif found in III class promoters. However, this motif is smaller in size and lesser expressed in II class promoters. 

As indicated above, in contrast to III class promoters, the promoters of II class contain no AT-rich component in their upstream region. The absence of this component could explain the difference in the form (size and value) of the electrostatic motif at  18 b.p. for the promoters of the different classes. What’s more, it suggests that the corresponding site A of the enzyme (Fig. 3) involved in binding of AT-rich promoter region is not engaged in electrostatic interactions with II class promoters thus emphasizing differential character of electrostatic contacts in primary complexes of T7 RNA polymerase with promoters of II and III classes in this part of promoter DNA.

As can be seen from Fig. 4 and Fig. 6, the most negatively charged element with the center at ~ -5 b.p. dominates in the electrostatic profiles of II class promoters. The presence of the electrostatic element in such an explicit form should be indicative of its role for promoter activity but mechanisms of its participation in promoter-polymerase complex formation are not quite clear. The element is located just at the boundary between “binding” and “initiation” regions. The involvement of the “initiation” region in promoter-polymerase interaction at late steps of complex-formation is considered to be unaffected by its electrostatic properties. Taking into account electrostatic nature of the element it is apparent that it should be involved in primary identification of II class promoters by RNA polymerase thus compensating for “poor” electrostatic element at -18 b.p. in their profiles. However, such an arrangement of the primary complex can complicate efforts for its further interconversion into the open “melted” complex. 

Promoter region corresponding to the electrostatic element at -5 b.p. is known to bind with the site C of the enzyme which as a whole is positively charged area (Fig. 3), although its total positive potential is not so well pronounced as in the case of the other DNA binding enzyme sites. Thus, electrostatic promoter-polymerase interactions are possible in this region of promoter DNA as well but the influence of these electrostatic contacts on the overall process of productive complex formation differs for the different promoter classes and is not clearly understandable by mechanisms.

Summarizing, the results obtained indicate quite different arrangement of electrostatic interactions in primary complexes of RNA polymerase with promoters of II and III classes.

Primary RNA polymerase binding based on electrostatic interactions can also play a structural role (Von Hippel, 2004). This binding serves to orient the protein properly with respect to the DNA region so as to facilitate specific aminoacid-nucleotide recognition within the grooves of the double helix DNA and provide further conformational changes of promoter-polymerase complex. The difference in the arrangement of primary complexes formed by the two electrostatic elements may affect the pathway and the rate of the overall process of the formation of the final active complex thus suggesting a plausible explanation for the difference in “strengths” and functional behavior of the promoters belonging to II or III classes.

Note although two promoters (φ2.5 and φ3.8) are located in class II region of the genome and assigned to this class by their physiological properties, however their electrostatic profiles are more similar in design to those of III class promoters (Fig. 3). Their negatively charged sites around -18 b.p. are of greater intensity than their electrostatic elements at -5 b.p. what is characteristic of III class promoters. The characteristic element at -5 b.p. is at all poorly expressed in their profiles what differ them from II class promoters. It is interesting that in accordance with literature data one of the promoters (namely φ3.8) behaves as III class promoters by its biochemical properties (Jolliffe, Carter, \& McAllister, 1982; McAllister et al., 1981), thus confirming that it is just DNA electrostatic characteristics that are important for promoter function and behavior.

Thus, the results obtained indicate that electrostatic patterns of T7 specific promoters can be characterized by some distinctive motifs which are specific for each promoter class. It should be noted that electrostatic profiles of the individual promoters belonging to the same class are similar but not strictly identical. Variations in some parameters of the same common motif can somewhat modulate its contribution in activity of the individual promoters but what is important that they do not change specificity of its spatial interaction with RNA polymerase thus keeping up the common specificity of the corresponding promoter class. By contrast, the difference in the distinctive motifs for promoters of different classes results in difference in their recognition by RNA polymerase thus differentiating their functional behaviour.

In summary, T7 RNA polymerase can identify its native promoters in T7 genome and differentiate them due to their electrostatic features.
\clearpage

\section{References}
\begin{enumerate}
\item Alexandrov, N. N., \& Mironov, A. A. (1990). Application of a new method of pattern recognition in DNA sequence analysis: a study of \textit{E. coli} promoters. \textit{Nucleic acids research, 18(7)}, 1847–52. Retrieved from http://www.pubmedcentral.nih.gov/\\
articlerender.fcgi?artid=330605\&tool=pmcentrez\&rendertype=abstract
\item Baker, N. A., Sept, D., Joseph, S., Holst, M. J., \& McCammon, J. A. (2001). Electrostatics of nanosystems: application to microtubules and the ribosome. \textit{Proceedings of the National Academy of Sciences of the United States of America, 98(18),} 10037–10041. doi:10.1073/pnas.181342398
\item Bénichou, O., Kafri, Y., Sheinman, M., \& Voituriez, R. (2009). Searching Fast for a Target on DNA without Falling to Traps. \textit{Physical Review Letters, 103(13),} 138102. doi:10.1103/PhysRevLett.103.138102
\item Cheetham, G. M., Jeruzalmi, D., \& Steitz, T. A. (1999). Structural basis for initiation of transcription from an RNA polymerase-promoter complex. \textit{Nature, 399(6731),} 80–3. doi:10.1038/19999
\item deHaseth, P. L., \& Helmann, J. D. (1995). Open complex formation by \textit{Escherichia coli} RNA polymerase: the mechanism of polymerase-induced strand separation of double helical DNA. \textit{Molecular microbiology, 16(5),} 817–24. Retrieved from http://www.ncbi.nlm.nih.gov/pubmed/7476180
\item Dunn, J. J., Studier, F. W., \& Gottesman, M. (1983). Complete nucleotide sequence of bacteriophage T7 DNA and the locations of T7 genetic elements. \textit{Journal of Molecular Biology, 166(4),} 477–535. doi:10.1016/S0022-2836(83)80282-4
\item Durniak, K., Bailey, S., \& Steitz, T. (2008). The Structure of a Transcribing T7 RNA Polymerase in Transition from Initiation to Elongation. \textit{Science (New York, NY), 322(5901)}, 553.
\item Golomb, M., \& Chamberlin, M. (1974). Characterization of T7-specific ribonucleic acid polymerase. IV. Resolution of the major in vitro transcripts by gel electrophoresis. \textit{The Journal of biological chemistry, 249(9), }2858–63. Retrieved from http://www.ncbi.nlm.nih.gov/pubmed/4828324
\item Gordon, L., Chervonenkis, A. Y., Gammerman, A. J., Shahmuradov, I. A., \& Solovyev, V. V. (2003). Sequence alignment kernel for recognition of promoter regions. \textit{Bioinformatics, 19(15),} 1964–1971. doi:10.1093/bioinformatics/btg265
\item Gunderson, S. I., Chapman, K. A., \& Burgess, R. R. (1987). Interactions of T7 RNA polymerase with T7 late promoters measured by footprinting with methidiumpropyl-EDTA-iron(II). \textit{Biochemistry, 26(6),} 1539–46. Retrieved from \\
http://www.ncbi.nlm.nih.gov/pubmed/3036203
\item Hess, B., Kutzner, C., van der Spoel, D., \& Lindahl, E. (2008). GROMACS 4: Algorithms for Highly Efficient, Load-Balanced, and Scalable Molecular Simulation. \textit{Journal of Chemical Theory and Computation, 4(3), }435–447. doi:10.1021/ct700301q
\item Hertz, G. Z., \& Stormo, G. D. (1996). \textit{Escherichia coli }promoter sequences: analysis and prediction. \textit{Methods in enzymology, 273, }30–42. Retrieved from http://www.ncbi.nlm.nih.gov/pubmed/8791597
\item Horton, P. B., \& Kanehisa, M. (1992). An assessment of neural network and statistical approaches for prediction of \textit{E. coli }promoter sites. \textit{Nucleic acids research, 20(16),} 4331–8. Retrieved from http://www.pubmedcentral.nih.gov/\\
articlerender.fcgi?artid=334144\&tool=pmcentrez\&rendertype=abstract
\item Huerta, A. M., \& Collado-Vides, J. (2003). Sigma70 promoters in \textit{Escherichia coli}: specific transcription in dense regions of overlapping promoter-like signals. \textit{Journal of molecular biology, 333(2),} 261–78. Retrieved from \\
http://www.ncbi.nlm.nih.gov/pubmed/14529615
\item Ikeda, R. A., \& Richardson, C. C. (1987). Interactions of a proteolytically nicked RNA polymerase of bacteriophage T7 with its promoter. \textit{The Journal of biological chemistry, 262(8), }3800–8. Retrieved from \\
http://www.ncbi.nlm.nih.gov/pubmed/3546320
\item Joho, K. E., Gross, L. B., McGraw, N. J., Raskin, C., \& McAllister, W. T. (1990). Identification of a region of the bacteriophage T3 and T7 RNA polymerases that determines promoter specificity. \textit{Journal of molecular biology, 215(1),} 31–9. doi:10.1016/S0022-2836(05)80092-0
\item Jolliffe, L. K., Carter, A. D., \& McAllister, W. T. (1982). Identification of a potential control region in bacteriophage T7 late promoters. \textit{Nature, 299(5884), }653–6. Retrieved from http://www.ncbi.nlm.nih.gov/pubmed/7121598
\item Jorgensen, W. L., \& Tirado-Rives, J. (1988). The OPLS potential functions for proteins, energy minimizations for crystals of cyclic peptides and crambin. \textit{Journal of the American Chemical Society, 110(6),} 1657–1666.
\item Kamzolova, S. G., Ivanova, N. N., \& Kamzalov, S. S., (1999). Long-range interactions in T2 DNA during its complex formation with RNA polymerase from \textit{E.coli}. \textit{Journal of Biological Physics, 24,} 157–161.
\item Kamzolova, S. G., Osypov, A. A., Dzhelyadin, T. R., Beskaravainy, P. M., \& Sorokin, A. A. (2006). Context-dependent effects of upstream A-tracts on promoter electrostatic properties and function. \textit{Proceedings of the Fifth International Conference on Bioinformatics of Genome Regulation and Structure, Vol 1 }(pp. 56–60).
\item Kamzolova, S. G., \& Postnikova, G. Y. B. (1981). Spin-Labelled nucleic acids. \textit{Quarterly reviews of biophysics, 14(2), }223–288. Retrieved from \\
http://www.ncbi.nlm.nih.gov/pubmed/6169107
\item Kamzolova, S. G., Sivozhelezov, V. S., Sorokin, A. A., Dzhelyadin, T. R., Ivanova, N. N., \& Polozov, R. V. (2000). RNA polymerase-promoter recognition. Specific features оf electrostatic potential of “early” Т4 phage DNA promoters. \textit{Journal of Biomolecular Structure \& Dynamics, 18(3),} 325–334. Retrieved from \\
http://www.ncbi.nlm.nih.gov/pubmed/11149509
\item Kamzolova, S. G., Sorokin, A. A., Beskaravainy, P. M., \& Osypov, A. A. (2006). Comparative analysis of electrostatic patterns for promoter and non promoter DNA in \textit{E.coli}. \textit{Bioinformatics of Genome Regulation and Structure II, }67–74.
\item Kamzolova, S. G., Sorokin, A. A., Dzhelyadin, T. R. D., Beskaravainy, P. M., \& Osypov, A. A. (2005). Electrostatic potentials of \textit{E.coli }genome DNA. \textit{Journal of Biomolecular Structure \& Dynamics, 23(3),} 341–346. Retrieved from \\
http://www.ncbi.nlm.nih.gov/entrez/query.fcgi?db=pubmed\&cmd=Retrieve\&\\
dopt=AbstractPlus\&list\_uids=16218758
\item Kamzolova, S. G., Sorokin, A. A., Osipov, A. A., \& Beskaravainy, P. M. (2009). Electrostatic Map of Bacteriophage T7 Genome. Comparative Analysis of Electrostatic Properties of sigma(70)-Specific T7 DNA Promoters Recognized by RNA-Polymerase of \textit{Escherichia coli}. \textit{Biofizika, 54(6),} 975-983.
\item Kassavetis, G. A., \& Chamberlin, M. J. (1979). Mapping of class II promoter sites utilized in vitro by T7-specific RNA polymerase on bacteriophage T7 DNA. \textit{Journal of virology, 29(1),} 196–208. Retrieved from http://www.pubmedcentral.nih.gov/\\
articlerender.fcgi?artid=353100\&tool=pmcentrez\&rendertype=abstract
\item Kiefer, F., Arnold, K., Künzli, M., Bordoli, L., \& Schwede, T. (2009). The SWISS-MODEL Repository and associated resources. \textit{Nucleic Acids Research, 37}(Database issue), D387–92. doi:10.1093/nar/gkn750
\item Leirmo, S., \& Gourse, R. L. (1991). Factor-independent activation of \textit{Escherichia coli} rRNA transcription. I. Kinetic analysis of the roles of the upstream activator region and supercoiling on transcription of the rrnB P1 promoter in vitro. \textit{Journal of molecular biology, 220(3),} 555–68. Retrieved from \\
http://www.ncbi.nlm.nih.gov/pubmed/1870123
\item Mahadevan, I., \& Ghosh, I. (1994). Analysis of \textit{E.coli} promoter structures using neural networks. \textit{Nucleic acids research, 22(11),} 2158–65. Retrieved from http://www.pubmedcentral.nih.gov/articlerender.fcgi?artid=308136\&tool=pmcentrez\&\\
rendertype=abstract
\item Margalit, H., Shapiro, B. A., Nussinov, R., Owens, J., \& Jernigan, R. L. (1988). Helix stability in prokaryotic promoter regions. \textit{Biochemistry, 27(14),} 5179–88. Retrieved from http://www.ncbi.nlm.nih.gov/pubmed/3167040
\item McAllister, W. T., \& Carter, A. D. (1980). Regulation of promoter selection by the bacteriophage T7 RNA polymerase in vitro. \textit{Nucleic acids research, 8(20),} 4821–37. Retrieved from \\
http://www.pubmedcentral.nih.gov/articlerender.fcgi?artid=324390\&tool=pmcentrez\\
\&rendertype=abstract
\item McAllister, W. T., Morris, C., Rosenberg, A. H., \& Studier, F. W. (1981). Utilization of bacteriophage T7 late promoters in recombinant plasmids during infection. \textit{Journal of molecular biology, 153(3),} 527–44. Retrieved from \\
http://www.ncbi.nlm.nih.gov/pubmed/7040687
\item McAllister, W. T., \& Wu, H. L. (1978). Regulation of transcription of the late genes of bacteriophage T7. \textit{Proceedings of the National Academy of Sciences of the United States of America, 75(2),} 804–8. Retrieved from http://www.pubmedcentral.nih.gov/\\
articlerender.fcgi?artid=411345\&tool=pmcentrez\&rendertype=abstract
\item McAllister, William T., \& McCarron, R. J. (1977). Hybridization of the in vitro products of bacteriophage T7 RNA polymerase to restriction fragments of T7 DNA. \textit{Virology, 82(2),} 288–298. doi:10.1016/0042-6822(77)90004-6
\item Muller, D. K., Martin, C. T., \& Coleman, J. E. (1989). T7 RNA polymerase interacts with its promoter from one side of the DNA helix. \textit{Biochemistry, 28(8),} 3306–13. Retrieved from http://www.ncbi.nlm.nih.gov/pubmed/2545254
\item Niles, E. G., \& Condit, R. C. (1975). Translational mapping of bacteriophage T7 RNAs synthesized In vitro by purified T7 RNA polymerase. \textit{Journal of Molecular Biology, 98(1),} 57–67. doi:10.1016/S0022-2836(75)80101-X
\item Osypov, A. A., Krutinin, G. G., \& Kamzolova, S. G. (2010). Deppdb - DNA electrostatic potential properties database: electrostatic properties of genome DNA. \textit{Journal of bioinformatics and computational biology, 8(3),} 413–425. Retrieved from http://www.ncbi.nlm.nih.gov/pubmed/20556853
\item Pedersen, A. G., \& Engelbrecht, J. (1995). Investigations of \textit{Escherichia coli} promoter sequences with artificial neural networks: new signals discovered upstream of the transcriptional startpoint. \textit{Proceedings of International Conference on Intelligent Systems for Molecular Biology, 3,} 292–9. Retrieved from \\
http://www.ncbi.nlm.nih.gov/pubmed/7584449
\item Polozov, R. V, Dzhelyadin, T. R., Sorokin, A. A., Ivanova, N. N., Sivozhelezov, V. S., \& Kamzolova, S. G. (1999). Electrostatic potentials of DNA. Comparative analysis of promoter and nonpromoter nucleotide sequences. \textit{Journal of Biomolecular Structure \& Dynamics, 16(6),} 1135 – 43. Retrieved from http://www.ncbi.nlm.nih.gov/\\
entrez/query.fcgi?db=pubmed\&cmd=Retrieve\&dopt=AbstractPlus\&list\_uids=10447198
\item Pérez-Martín, J., Rojo, F., \& De Lorenzo, V. (1994). Promoters responsive to DNA bending: a common theme in prokaryotic gene expression. \textit{Microbiological reviews, 58(2),} 268–90. Retrieved from http://www.pubmedcentral.nih.gov/articlerender.fcgi?\\
artid=372964\&tool=pmcentrez\&rendertype=abstract
\item Sheinman, M., Bénichou, O., Kafri, Y., \& Voituriez, R. (2012). Classes of fast and specific search mechanisms for proteins on DNA. \textit{Reports on Progress in Physics, 75(2),} 026601. doi:10.1088/0034-4885/75/2/026601
\item Slutsky, M., \& Mirny, L. A. (2004). Kinetics of Protein-DNA Interaction: Facilitated Target Location in Sequence-Dependent Potential. \textit{Biophysical Journal, 87(6),} 4021–4035. doi:10.1529/biophysj.104.050765
\item Sorokin, A. A., Osipov, A. A., Beskaravaĭnyĭ, P. M., \& Kamzolova, S. G. (2007). Analysis of the distribution of the nucleotide sequence and electrostatic potential of the \textit{Escherichia coli} genome. \textit{Biofizika, 52(2),} 223–7. Retrieved from \\
http://www.ncbi.nlm.nih.gov/pubmed/17477048
\item Sorokin, Anatoly A. (2001). Functional analysis of \textit{E.coli} promoter sequences. New promoter determinants. (Doctoral dissertation, Institute of Theoretical and Experimental Biophysics RAS, 2001)
\item Sorokin, A. A., Dzhelyadin, T. R., Ivanova, N. N., Polozov, R. V, \& Kamzolova, S. G. (2001). The quest for new forms of promoter determinants. Relationship of promoter nucleotide sequences to their electrostatic potential distribution. \textit{Journal of Biomolecular Structure \& Dynamics, 18(6), }1020. Retrieved from http://www.jbsdonline.com/The-quest-for-new-forms-of-promoter-determinants- \\ 
Relationship -of-promoter-nucleotide -sequences-to-their-electrostatic-potential- \\
distribution-p10233.html
\item Sorokin, A. A., Osypov, A. A., Dzhelyadin, T. R., Beskaravainy, P. M., \& Kamzolova, S. G. (2006). Electrostatic properties of promoter recognized by \textit{E.coli} RNA polymerase Esigma70. \textit{Journal of bioinformatics and computational biology, 4(2),} 455 – 67. Retrieved from http://www.ncbi.nlm.nih.gov/pubmed/16819795
\item Studier, F. W., \& Rosenberg, A. H. (1981). Genetic and physical mapping of the late region of bacteriophage T7 DNA by use of cloned fragments of T7 DNA. \textit{Journal of Molecular Biology, 153(3),} 503–525. doi:10.1016/0022-2836(81)90405-8
\item Travers, A. A. (1989). DNA conformation and protein binding. \textit{Annual review of biochemistry, 58,} 427–52. doi:10.1146/annurev.bi.58.070189.002235
\item Vanet, A., Marsan, L., \& Sagot, M.-F. (1999). Promoter sequences and algorithmical methods for identifying them. \textit{Research in Microbiology, 150(9-10),} 779–799. doi:10.1016/S0923-2508(99)00115-1
\item Von Hippel, P. H. (2004). Biochemistry. Completing the view of transcriptional regulation. \textit{Science, 305(5682),} 350–352. doi:10.1126/science.1101270
\item Yada, T., Nakao, M., Totoki, Y., \& Nakai, K. (1999). Modeling and predicting transcriptional units of \textit{Escherichia coli }genes using hidden Markov models. \textit{Bioinformatics, 15(12), }987–993. doi:10.1093/bioinformatics/15.12.987

\end{enumerate}

\clearpage

\begin{table}[htbp]
  \caption{Promoters of T7 bacteriophage recognized by T7 RNA polymerase.
23-membered consensus sequence is shown on the 1st line. In the promoters this sequence is emphasized by spaces; transcription start site (TSS) is marked by capital letter; nucleotides which differ from consensus are shown in bold; nucleotides identical in all promoters are undelined.
}
  \label{tab:myfirsttable}
  \centering
  \begin{tabular}{ | c | c | c |} 
\hline
Promoter & Position of TSS (b.p.) & Nucleotide sequence \\
\hline
\multicolumn{2}{|c|}{consensus sequence}	& -17 taatacgactcactataGggaga +6\\
\hline
φOL & 405 & gtctttat taatacaactcactataAggaga gaca \\
\hline
\multicolumn{3}{| c |}{«Early» promoters (subgroup of class II promoters)} \\
\hline
φ1.1A & 5848 & cgccaaat caatacgactcactataGaggga caaa \\
φ1.1B & 5923 & cttccggt taatacgactcactataGgagaa cctt \\
φ1.3 & 6409 & actggaag taatacgactcagtataGggaca atgc \\
\hline
\multicolumn{3}{| c |}{Class II promoters} \\
\hline
φ1.5 & 7778 & ttaactgg taatacgactcactaaaGgaggt acac \\
φ1.6 & 7895 & gtcacgct taatacgactcactaaaGgagac acta \\
φ2.5 & 	9107 & 	caccgaag taatacgactcactattAgggaa gact \\
φ3.8 & 	11180 & 	tggataat taattgaactcactaaaGggaga ccac \\
φ4c	 & 12671 & 	gactgaga caatccgactcactaaaGagaga gatt \\
φ4.3 & 	13341 & 	tcccattc taatacgactcactaaaGgagac acac \\
φ4.7 & 	13915 & 	catgaata ctattcgactcactataGgagat atta \\
\hline
\multicolumn{3}{|c|}{Class III promoters} \\
\hline
φ6.5	 & 18545 & 	ccctaaat taatacgactcactataGggaga tagg \\
φ9	 & 21865 & 	cgggaatt taatacgactcactataGggaga cctc \\
φ10 & 	22904 & 	ttcgaaat taatacgactcactataGggaga ccac \\
φ13	 & 27274 & 	ctcgaaat taatacgactcactataGggaga acaa \\
φ17	 & 34566 & 	gtaggaaa taatacgactcactataGggaga ggcg \\
\hline
φOR	 & 39229 & 	cgataaat taatacgactcactataGggaga ggag \\
\hline
\end{tabular}
\end{table}

\begin{figure}[htbp]
\begin{center}
\includegraphics[scale=0.6]{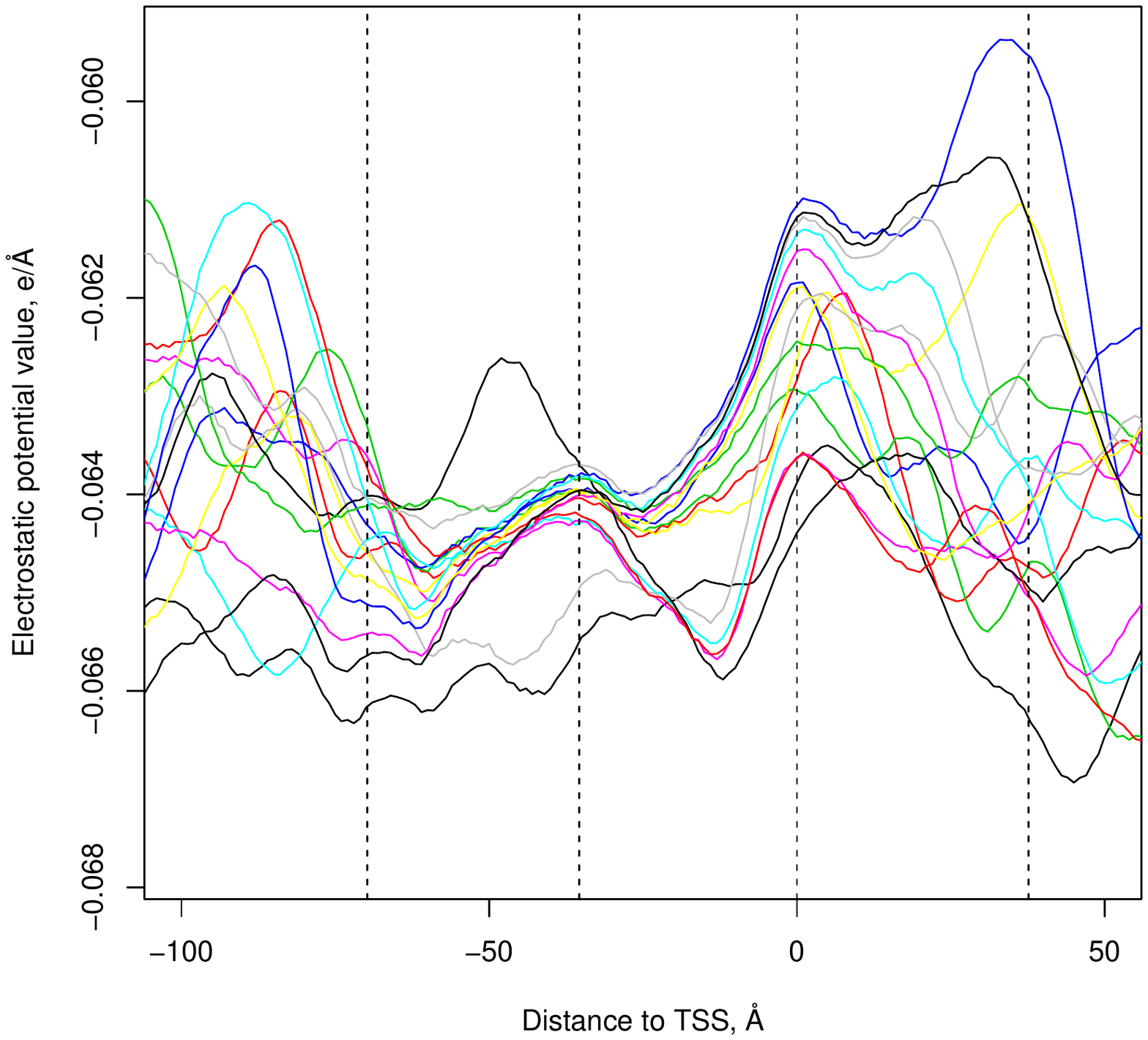}
\caption{Distribution of electrostatic potential profiles around of all T7 specific promoters.}
\label{fig:map}
\end{center}
\end{figure}

\begin{figure}[htbp]
\begin{center}
\includegraphics[scale=0.6]{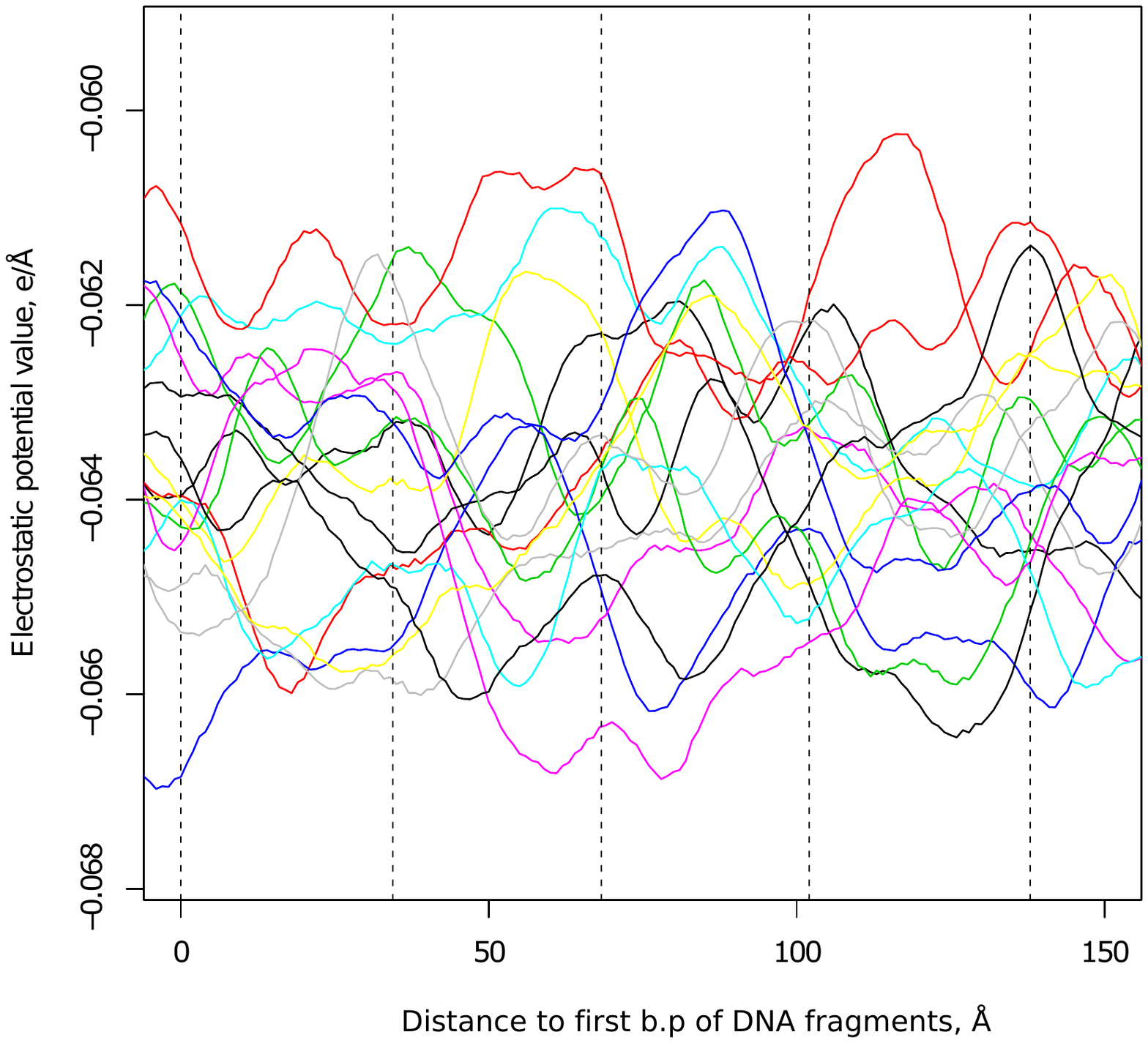}
\caption{Distribution of electrostatic potential profiles around T7 DNA at randomly selected T7DNA sequences.}
\label{fig:map}
\end{center}
\end{figure}

\begin{figure}[htbp]
\begin{center}
\includegraphics[scale=0.6]{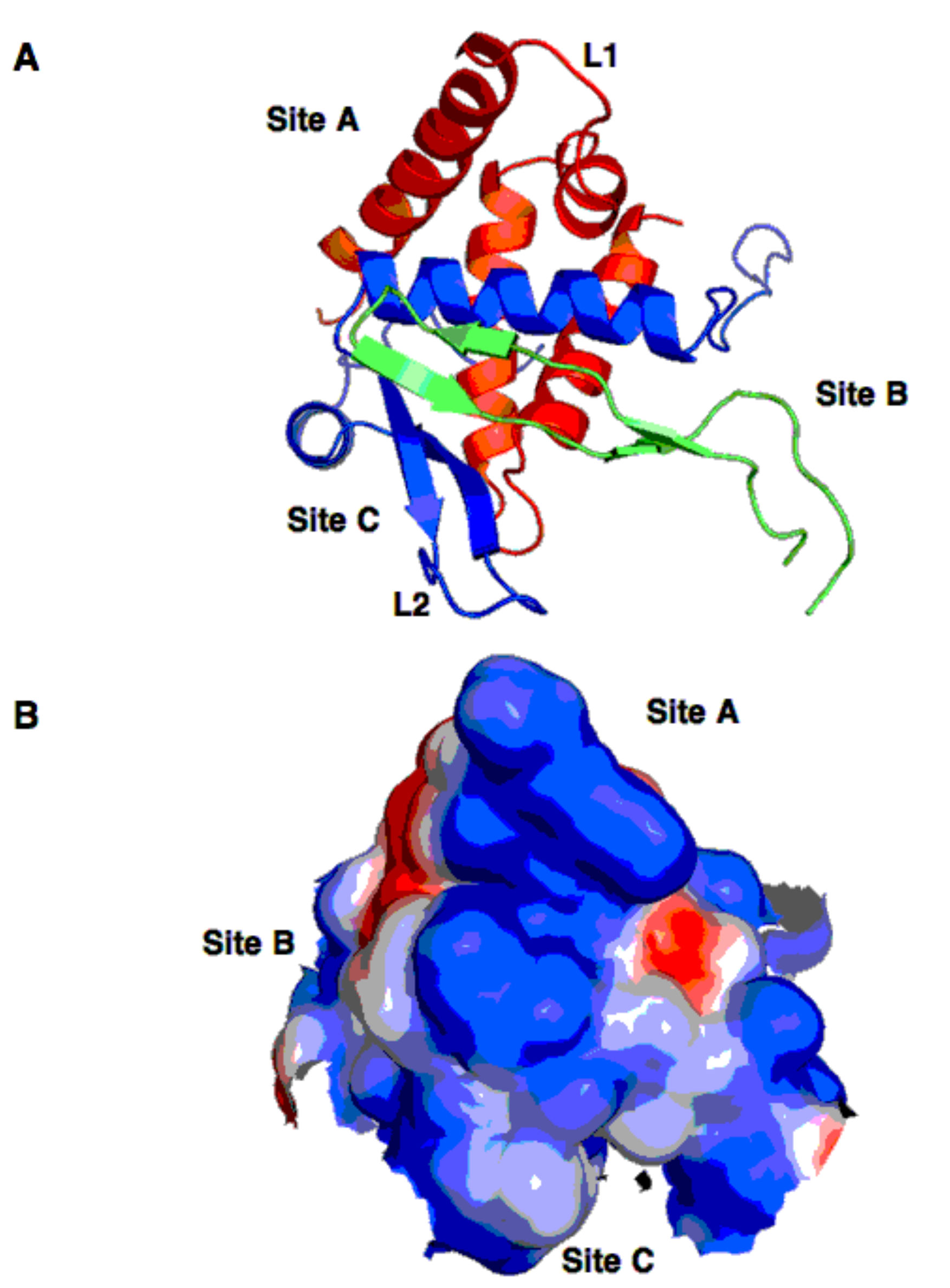}
\caption{Sturucture and electrostatic properties of promoter-binding domain (PBD) of T7 RNA polymerase. A) cartoon representation of three sub-domains constituting PBD: in red site A (residues 72-150), in green site B (residues 739-770), in blue site C (residues 191-267). Two loop responsible for features on the electrostatic profile are designated. B) distribution of electostatic potential on the solvent accessible surgace of T7 RNA polymerase PBD. Three crests are formed by (from top to bottom) site A upper loop (L1 on the panel A), site B, and site C bottom loop (L3 on the panel A). }
\label{fig:map}
\end{center}
\end{figure}

\begin{figure}[htbp]
\begin{center}
\includegraphics[scale=0.6]{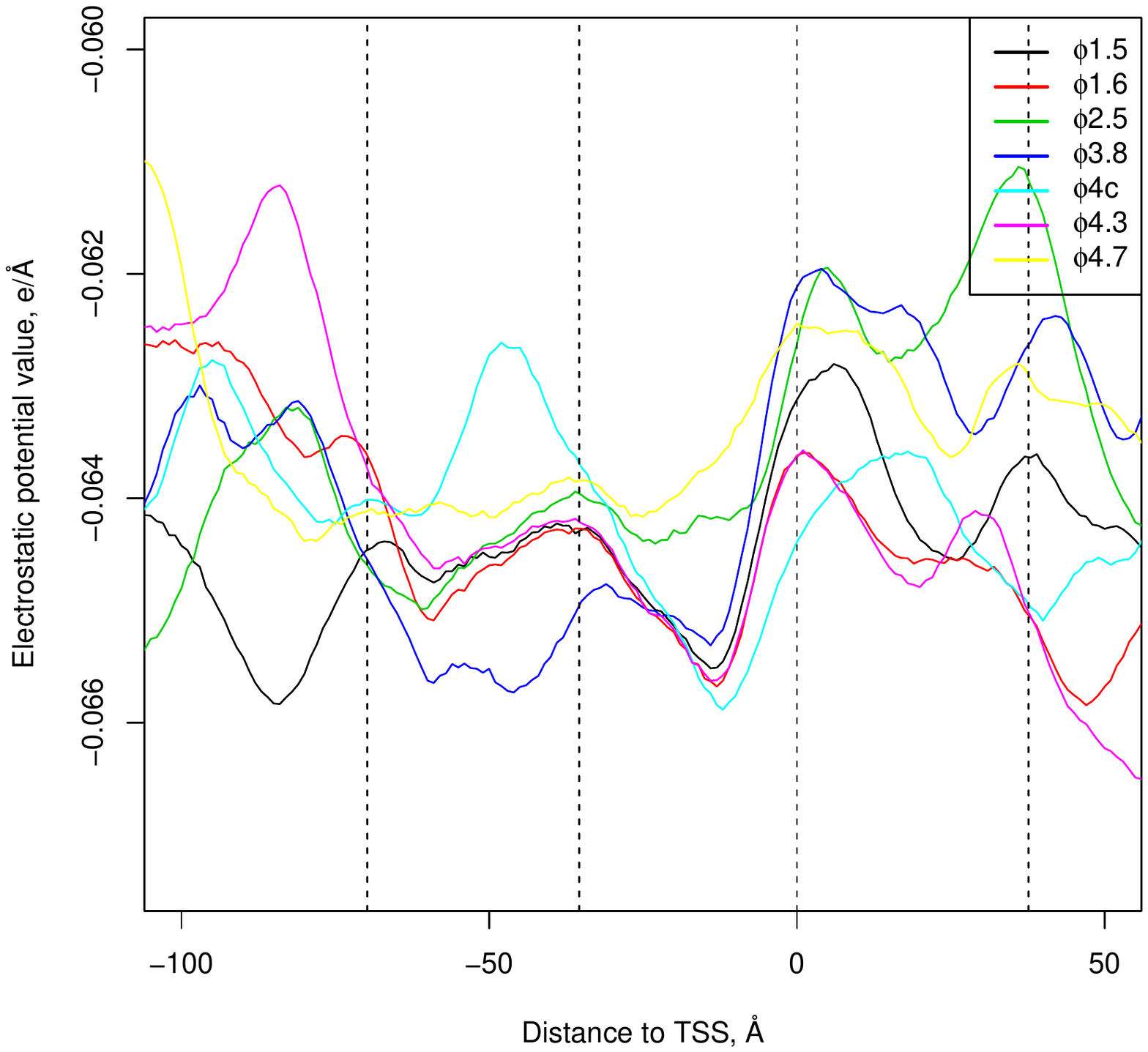}
\caption{Distribution of electrostatic potential around II class T7 promoters.}
\label{fig:map}
\end{center}
\end{figure}

\begin{figure}[htbp]
\begin{center}
\includegraphics[scale=0.6]{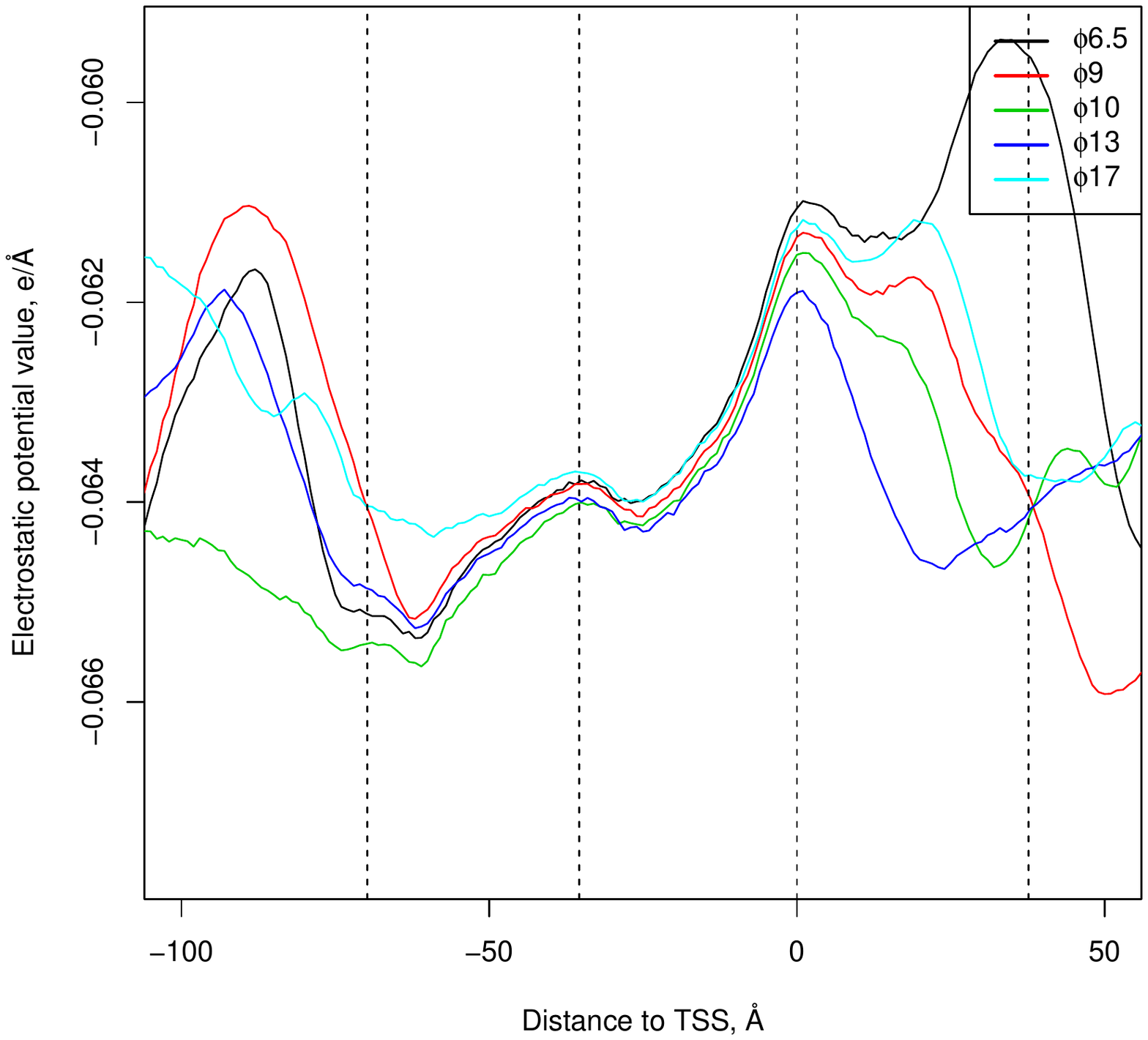}
\caption{Distribution of electrostatic potential around III class T7 promoters.}
\label{fig:map}
\end{center}
\end{figure}

\begin{figure}[htbp]
\begin{center}
\includegraphics[scale=0.6]{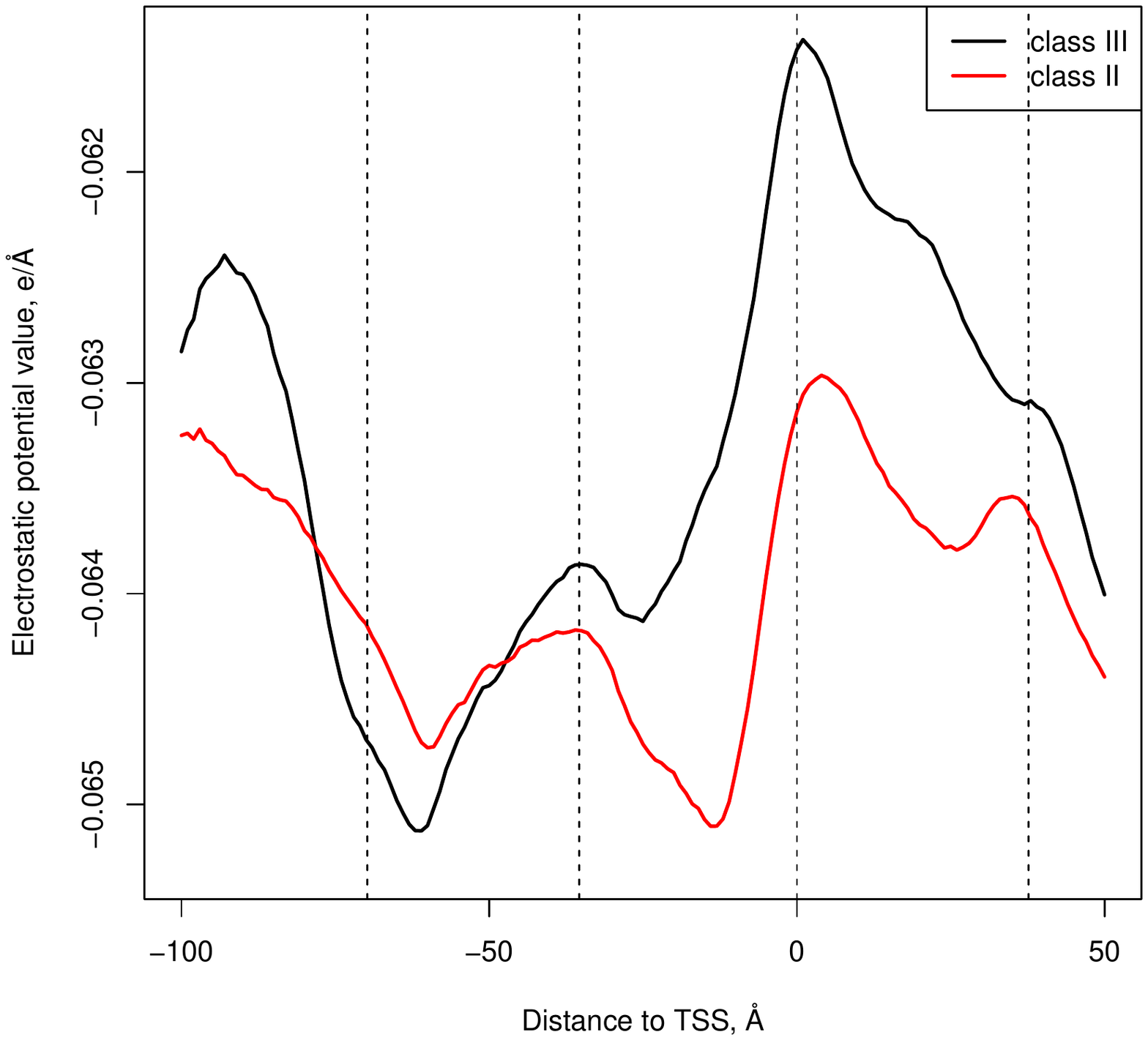}
\caption{The averaged profiles of electrostatic potential distribution around T7 promoters belonging to different classes.}
\label{fig:map}
\end{center}
\end{figure}

\clearpage

\appendix
\section{Method of calculation of the electrostatic potential around long DNA sequences}

Full-atom model of DNA molecule was used with atom coordinates taken from (Aoki et al 1988) and charges assigned according to (Zhurkin et al 1980). The helix geometry of DNA was reconstructed on the basis of  nucleotide sequence according to (Babcock, Olson 1994). To keep the main axis of DNA double helix straight, only rise and twist structural parameters of dinucleotide step (Dickerson 1989) were set to non-zero values. Values of rise and twist structural parameters corresponding to 16 individual dinucleotides were taken from (Ponomarenko et al. 1997). 
Calculations of electrostatic potential $\phi(\vec{r})$  of DNA were carried out on the surface of the cylinder coaxial to the DNA helix with radius 15~\AA~ in accordance with Coulomb’s law: 

 \begin{equation}
\phi(\vec{r})=\sum_{i}{\frac{q_{i}}{\epsilon(\vec{r})\left|\vec{r}-\vec{r_{i}}\right|}}
\end{equation}

where $q_{i}$ — partial charge of i-th atome of the DNA molecule; $\vec{r_{i}}$ — position of the  i-th atom;  $\vec{r}$— position of observation point; $\epsilon(\vec{r})$  — position dependent dielectric constant. 

To take into account counterion condensation on the sugar-phosphate backbone of the DNA partial charges of O1 and O2 atoms of phosphate groups were decreased by 50\%. Dielectric constant is chosen to be proportional to the distance from the charge to the point of consideration.

The choice of cylinder radius, level of charge screening and form of the dielectric constant function were obtained by optimization of agreement between Coulomb-based potential and the pontential calculated by solution of Poisson-Boltzman equation around short DNA fragments (Polozov et al. 1999).

The cut-off distance of 50 b.p was introduced to improve performance of calculation procedures and make tractable the analysis of electrostatic potential of the whole genome length DNA. This cut-off means that only 100 bp fragment around point of consideration can influence the electrostatic potential value.

For ease of interpretation and numerical analysis, 2D electrostatic potential map is represented as the 1D profile, where the profile magnitude at each coordinate along the DNA axis is an average of the electrostatic potential values on the circle perpendicular to the helix at that coordinate.

\subsection{References}
\begin{enumerate}
\item Aoki, K., Arnott, S., Chandrasekaran, R., Jeffrey, G. A., Moras, D., \& Neidle, S. (1988). \textit{Crystallographic and Structural Data II / Kristallographische und strukturelle Daten II.} Springer.
\item Zhurkin, V. B., Poltev, V. I., \& Florent'ev, V. L. (1980). Atom-atomic potential functions for conformational calculations of nucleic acids. \textit{Molekuliarnaia biologiia, 14(5),} 1116–1130.
\item Babcock, M. S., \& Olson, W. K. (1994). The effect of mathematics and coordinate system on comparability and “dependencies” of nucleic acid structure parameters. \textit{Journal of Molecular Biology, 237(1),} 98–124. doi:10.1006/jmbi.1994.1212
\item Ponomarenko, M. P., Ponomarenko, I. V., Kel, A. E., Kolchanov, N. A., Karas, H., Wingender, E., \& Sklenar, H. (1997). Computer analysis of conformational features of the eukaryotic TATA-box DNA promoters. \textit{Molekuliarnaia biologiia, 31(4),} 623–630.
\item Polozov, R. V., Dzhelyadin, T. R., Sorokin, A. A., Ivanova, N. N., Sivozhelezov, V. S., \& Kamzolova, S. G. (1999). Electrostatic potentials of DNA. Comparative analysis of promoter and nonpromoter nucleotide sequences. \textit{Journal of Biomolecular Structure \& Dynamics, 16(6),} 1135–1143.
\item Dickerson, R. E. (1989). Definitions and nomenclature of nucleic acid structure parameters. \textit{The EMBO Journal, 8(1),} 1–4.
\end{enumerate}
\end{document}